\documentclass[12pt,letter]{article}

 \usepackage{epsfig}
\usepackage{amsfonts}
\usepackage{amsmath}
\usepackage{amsthm}
\usepackage{amssymb}
\usepackage{algorithm}
\usepackage[noend]{algpseudocode}
\usepackage[hyphens]{url}
\usepackage{hyperref}
\usepackage{graphicx}
\usepackage{subfigure}
\usepackage{xspace}
\usepackage{units}
\usepackage[margin=1in]{geometry}

 \usepackage{array}
 \usepackage{multirow}
\usepackage{soul}
\usepackage{color}
\usepackage[table]{xcolor}
\usepackage{hhline}
\usepackage{amsmath}
\usepackage{amssymb}
\usepackage{graphicx}
\usepackage{enumitem}
\usepackage{booktabs}
\usepackage{comment}
\usepackage{supertabular}
\usepackage{xparse}
\usepackage{tikz}
\usepackage{colortbl}
\usepackage{soul}
\usetikzlibrary{calc,matrix,patterns,shadings,backgrounds}
\pgfdeclarelayer{myback}
\pgfsetlayers{myback,background,main}
\tikzset{myfillcolor/.style = {draw,fill=#1}}%

\definecolor{Gray}{gray}{0.9}
\NewDocumentCommand{\fillpattern}{O{north west lines} O{blue!50} m m}{%
	\draw[pattern=#1, pattern color=#2] (#3.north west)rectangle (#4.south east);
}

\begin{document}

\title{\bf Tempus Fugit: The Impact of Time in \\
 Knowledge Mobilization Networks}

\author{
Amir Afrasiabi Rad\footnotemark[1]
\and
Paola Flocchini\footnotemark[1]
\and
Joanne Gaudet\footnotemark[2]
 }

\def\thefootnote{\fnsymbol{footnote}}
\footnotetext[1]{School of Electrical Engineering and Computer Science, University of Ottawa, Canada.  a.afrasiabi@uOttawa.ca}
\footnotetext[2]{School of Computer Science, Carleton University, Canada.  flocchin@site.uottawa.ca}
\footnotetext[3]{ Alpen Path Solutions Inc. Ottawa, Canada.  jgaudet@magma.ca}

 \date{}

\maketitle

\begin{abstract}

The temporal component  of social networks   is often neglected in their analysis, 
and statistical measures are    typically performed on a ``static" representation of the network. As a result, 
measures of importance (like betweenness centrality) cannot reveal any temporal role of the entities involved.
Our goal is to start filling this limitation  by  proposing  a form of temporal betweenness measure, 
 and by  using it  to analyse a knowledge mobilization network. 
We show that  this measure, which  takes time explicitly into account, allows us to detect centrality 
roles that were completely hidden in the classical  statistical analysis.
In particular, we uncover nodes  whose  static centrality was considered negligible, 
but whose temporal role is instead  important  to  accelerate  mobilization flow in the network.
We also observe the reverse behaviour by detecting nodes with high static centrality, 
whose role as temporal bridges is instead very low.
By revealing important temporal roles, this study
is a first step towards a better understanding of the impact of time in
 social networks, and opens the road to further investigation.

  \end{abstract}

\maketitle 
\section{Introduction}

Highly dynamic networks are networks where connectivity changes in time and connection patterns display possibly complex dynamics. Such networks
are more and more pervasive in everyday life and the study of their properties is the object of extensive investigation 
 in a wide range  of very different contexts. 
 Some of these contexts are typically studied in computer science, such as wireless, adhoc networks, transportation, vehicular networks, satellites, military and robotic networks (e.g., see \cite{CFMS10,CFMS12,JLW07,KLO10,LW09b,MiCS14}); while others belong to totally different disciplines. This is the case for example, of   the nervous system, livestock trade, epidemiological networks, and multiple forms of social networks  (e.g., see \cite{KLHC13,LSS13,MBA12,QCL11,SVMM14,HS12}). Clearly, while being different in many ways, these domains display common features; {\em time-varying graphs} (TVGs) represent a model that formalizes highly dynamic networks encompassing the above contexts into a unique framework, and   emphasizes their temporal nature \cite{CFQS12}.  

{\em Knowledge Mobilization} (KM) refers to the use of knowledge towards the achievement of goals \cite{gaudet2013takes}. Scientists, for example, use published papers to produce new knowledge in further publications to reach professional goals. In contrast, patient groups can use scientific knowledge to help foster change in patient practices, and corporations can use scientific knowledge to reach financial goals. Recently, researchers have started to analyse knowledge mobilization networks (KMN) using a social network analysis (SNA) approach  (e.g.,  see \cite{Binz2014,Boland2012,chan2006synergy,Eppler2001,klenk2010quantifying}). In particular, \cite{gaudet2014PP} proposed a novel approach where a heterogeneous network composed of a main class of actors subdivided into three sub-types (individual human and non-human actors, organizational actors, and non-human mobilization actors) associated according to one relation, knowledge mobilization (a Mobilization-Network approach). Data covered a seven-year period with static networks for each year. The mobilization network was analysed using classical SNA measures (e.g., node centrality measures, path length, density) to produce understanding for KM using insights from network structure and actor roles \cite{gaudet2014PP}.
 
The KM SNA studies mentioned above, however, lack a fundamental component: in fact, their analysis is based on a static representation of KM networks, incapable of sufficiently accounting for the time of appearance and disappearance of relations between actors beyond static longitudinal analysis. Indeed, incorporating the temporal component into analysis is a challenging task, but it is undoubtedly a critical one, because time is an essential feature of these networks. Temporal analysis of dynamic graphs is in fact an important and extensively studied area of research (e.g., see  \cite{GVOM13,Kim2012,KosKW08,Kostakos09,santoro2011time,Tang2009,TBK07}), but there is still much to be discovered.
 In particular, most temporal studies simply consider network dynamics in successive static snapshots thus capturing only 
 a very partial temporal component by observing how static parameters evolve in time while the network changes. 
 Moreover, very little work has been dedicated to empirically evaluating the usefulness of   metrics in time 
  (e.g., see \cite{ACPQS11,kossinets2006empirical}).
 
In this paper, we represent KMN by TVGs and we propose to analyse them in a truly temporal setting. We provide, for the first time on a real data set, an empirical indication of the effectiveness of a temporal betweenness measure specifically designed for TVGs. In particular, we focus on data extracted from \cite{gaudet2014PP}, here referred to as {\em Knowledge-Net}.
We first consider static snapshots of Knowledge-Net corresponding to the seven years of its existence, and by studying the classical centrality measures in those time intervals, we provide rudimentary indications of the networks' temporal behaviour.
 To gain a finer temporal understanding, we then concentrate on {\em temporal betweenness} following a totally different approach. Instead of simply observing the static network over consecutive time intervals, we focus on the TVG that represent Knowledge-Net and we compute a form of betweenness that explicitly and globally takes time into account. We compare the temporal results that we obtain with classical static betweenness measures to gain insights into the impact that time has on the network structure and actor roles. We notice that, while many actors maintain the same role in static and dynamic analysis, some display striking differences. In particular, we observe the emergence of important actors that remained invisible in static analysis, and we advance explanations for these. Results show that the form of temporal betweenness we apply is effective at highlighting the role of nodes whose importance has a temporal nature (e.g., nodes that contribute to mobilization acceleration). This research opens the road to the study of other temporal measures designed for TVGs. 

\section{Knowledge-Net}

\subsection{Data description}

{\em Knowledge-Net}  is an heterogeneous network where nodes represent human and non-human actors (researchers, projects, conference venues, papers, presentations, laboratories), and edges represent knowledge mobilization between two actors. The network was collected for a period of seven years \cite{gaudet2014PP}. Once an entity or a connection is created, it remains in the system for the for entire period of the analysis.

Table \ref{tbl:actorType} provides a description of the {\em Knowledge-Net}   dataset. The dataset consists of 366 vertices and 750 edges in 2011. The number of entities and connections vary over times starting from only 10 vertices and 14 edges in 2005 and accumulating to the final network year in 2011. 
{\em Knowledge-Net}  is mainly comprised of non-human actors, 272 in total 
(non-human mobilization actors, NHMA, 
non-human individual actors, NHIA, 
and organizational actors, OA), in relation with 94 human actors (HA). 
Human actors include principle investigators (PI), highly qualified personnel (HQP) and collaborators (CO). It is through mobilization actors (NHMA) that individual, organizational actors and mobilization actors associate and mobilize knowledge to reach goals. For example, scientists mobilize knowledge through articles where not all contributing authors might be in relation with all other authors, yet all relate with the publication \cite{gaudet2014PP}. These non-human mobilization actors make up the bulk of the network including conference venues, presentations (invited oral, non-invited oral and poster), articles, journals, laboratories, research projects, websites, and theses.

\begin{table}
	\renewcommand{\arraystretch}{1.3}
	\caption{{\em Knowledge-Net} data set with characteristics of actors and their roles at different times}
	\label{tbl:actorType}
	\begin{center}
		\centering
		
\begin{tabular}{|m{12mm}|m{18mm}|m{18mm}|m{18mm}|m{22mm}|}
\hline
Start & Duration & {\#Nodes} &  {\#Edges} & Granularity \\
\hline
{2005} & {7 Years} & 366   & 750   &   {1 Year}\\
\hline
\end{tabular}

\begin{tikzpicture}
\matrix (m)[matrix of nodes, minimum width =11cm, ampersand replacement =\&, nodes in empty cells]
{
	   \\
};

\begin{pgfonlayer}{myback}
\fillpattern{m-1-1}{m-1-1}
\end{pgfonlayer}
\end{tikzpicture}

\begin{tabular}{|m{12mm}|m{8mm}|m{8mm}|m{8mm}|m{8mm}|m{8mm}|m{8mm}|m{8mm}|}
			\hline
			Actor Type &  2005 & 2006 & 2007 & 2008 & 2009 & 2010 & 2011 \\
			\hline
			HIA & 3 & 22 & 27  & 46 & 51 &76 & 94\\
			\hline
			NHIA &  0 & 3 & 6 & 9 & 9 & 9 & 15  \\
			\hline
			NHMA &  7 & 25 & 43 & 87 & 132 & 194 & 248 \\
			\hline
		         OA &  0 & 5 & 5 & 9 & 9 & 9 &   2\\
			\hline
			Total & 10 & 55 & 81 & 151 & 201 & 288 & 366\\
			\hline
		\end{tabular}

\end{center}
\end{table}

Classical statistical parameters have been calculated for Knowledge-Net, representing it as a static graph where the time of appearance of nodes and edges did not hold any particular meaning. In doing so, several interesting observations were made regarding the centrality of certain nodes as knowledge mobilizers and the presence of communities  \cite{gaudet2014PP}. 
In particular, all actor types increased in number over the 7 years indicating a rise in new mobilization relations over time. Although non-human individual actor absolute numbers remained small (ranging from 3 in 2006 to 15 in 2011), these actors were critical to making visible tacit (non-codified) knowledge mobilization from around the world (mostly laboratory material sharing, including from organizations and universities in the USA, from Norway, and from Canadian universities). Finally, embedded in human individual actor counts were individuals that the laboratory acknowledged in peer-reviewed papers, thus making further tacit and explicit knowledge mobilization visible.

\subsection{Analysis of consecutive snapshots}

To  provide more clear statistics on the Knowledge-Net dataset and a ground for better understanding of temporal metrics, we first     calculated classical statistical measures (e.g., node centrality measures, path length, density)
on seven static graphs, corresponding to the seven years of study. The average for each value for the graphs is calculated to represent a benchmark on how the rank for each node is compared to others. 

\begin{table}
	\renewcommand{\arraystretch}{1.3}
	\caption{Some static statistical parameters calculated for successive snapshots}
	\label{tbl:snapShot}
	\begin{center}
		\centering
		\begin{tabular}{|c|c|c|c|c|c|c|c|}
			\hline
			\ & 2005 & 2006 & 2007 & 2008 & 2009 & 2010 & 2011 \\
			\hline
			Ave. Degree & 1.40 & 1.32 & 1.63 & 1.84  & 1.98 & 2.02 & 2.04\\
			\hline
			Diameter & 4 & 5 & 5 & 6 & 6 & 6 & 6\\
			\hline
			Density & 0.31 & 0.04  & 0.04 & 0.02 & 0.02 & 0.01 & 0.01\\
			\hline
			\#Communities & 4 & 3 & 6 & 8 & 8 & 15 & 12\\
			\hline
			Modularity & 0.17 & 0.52 & 0.46 & 0.47 & 0.46 & 0.54 & 0.54\\
			\hline
			Ave. Clustering Coefficient & 0.41 & 0.06 & 0.21 & 0.22 & 0.20 & 0.24 & 0.23\\
			\hline
			Ave. Path Length & 2.04 & 3.04 & 3.06 & 3.26 & 3.34 & 3.46 & 3.50\\
			\hline
			Ave. Normalized Closeness & 0.51 & 0.33 & 0.33 & 0.31 & 0.30 & 0.29 & 0.29\\
			\hline
			Ave. Eccentricity & 3.10 & 4.41 & 4.40 & 4.70 & 4.80 & 4.83 & 4.83\\
			\hline
			Ave. Betweenness & 4.70 & 58.36 & 83.53 & 169.70 & 234.89 & 354.23 & 456.18\\
			\hline
			Ave. Normalized Betweenness & 0.13 & 0.03 & 0.02 & 0.01 & 0.01 & $\approx$ 0 & $\approx$ 0\\
			\hline
			Ave. Page Rank & 0.10 & 0.01 & 0.01 & $\approx$ 0 & $\approx$ 0 & $\approx$ 0 & $\approx$ 0\\
			\hline
			Ave. Eigenvector & 0.52 & 0.19 & 0.15 & 0.10 & 0.09 & 0.07 & 0.05\\
			\hline
		\end{tabular}
	\end{center}
\end{table}

The statistical data presented in Table \ref{tbl:snapShot} provides valuable information about the graph. The steady decrease in the (normalized) centrality values confirms that the network growth is not symmetric, so the centrality values have long tails. The low value of normalized betweenness, along with the low values for density, confirms that the graph is coupled in a way that there are a great number of shortest paths between any two arbitrary vertices in the graph.
This caused the betweenness for most vertices to  be similar and quite low when compared to the ones of nodes with the highest betweenness. Low average path length is a sign that the network presents small world characteristics and the knowledge mobilization to the whole network is expected to be conducted only in a few hops. Meanwhile, the decreasing graph density along with the increasing average degree represent the slow growth in the number of edges compared to the number of nodes. Escalation in the number of communities with increase in graph modularity metrics shows that the knowledge mobilization actors tend to form communities as time progresses. As the normalized average betweenness decreases steadily, it can be concluded that a few vertices at each community play the role of mediators and create the link between communities. 

Apart from these general observations, a static analysis of consecutive snapshots, does not provide deep temporal understanding. For example, it does not reflect which entities engage in knowledge mobilization in a timely fashion, e.g. by facilitating fast mobilization, or slowing mobilization flow.

To tackle some of these questions, we represent {\em Knowledge-Net} as a TVG and we propose to study it by employing  a form of temporal betweenness that makes use of time in an explicit manner.

\section{Time-Varying Graphs}

\subsection{Definition} 
Time-varying Graphs are graphs whose structure varies over time.
Following \cite{CFQS12}, a time-varying graph ({\em TVG}) is defined as  a quintuple ${\cal G} = {(V,E,{\cal T},\rho,\zeta)}$, where $V$ is a finite set of nodes; $E \subseteq V \times V$ is a finite set edges. The graph is considered within a finite time span $\mathcal{T}\subseteq\mathbb{T}$, called lifetime of the system.
$\rho: E \times {\cal T} \to \{0,1\}$ is the edge presence function, which indicates whether a given edge is available at a given time; $\zeta: E \times {\cal T} \to {\mathbb T}$, is the latency function, which indicates the time it takes to cross a given edge if starting at a given date. The model may, of course, be extended by defining the vertex presence function $(\psi:V\times\mathcal{T}\rightarrow\{0,1\})$, and vertex latency function $(\phi:V\times\mathcal{T}\rightarrow\{0,1\})$.
The footprint of ${\cal G}$ is a static graph composed by the union of all nodes and edges ever appearing during the lifetime $\mathbb{T}$.

\begin{figure}
\centering
\includegraphics[width=0.6\linewidth]{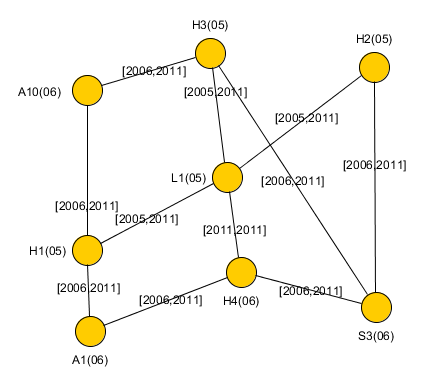}
\caption{A small portion of Knowledge-Net represented as a TVG.}
\label{fig:CasTVGViz}
\end{figure}

When representing {\em Knowledge-Net} as a TVG, we notice that the latency $\zeta$ is always zero, as an edge represents a relationship and its creation does not involve any delay; moreover, edges and nodes exist from their creation (their birth-date) to the end of the system lifetime. Let {\em birth-date}$(e)$ denote the year when edge $e$ is created. An example of a small portion of {\em Knowledge-Net} represented as a TVG is given in Figure \ref{fig:CasTVGViz}. 

\subsection{Journeys} 

A journey $\mathcal{J}$ in a TVG $\mathcal{G}$ is a temporal walk
defined as a sequence of ordered pairs $\{(e_1,t_1),(e_2,t_2)$,...,$(e_k,t_k)\}$, such that $\{e_1,e_2,...,e_k\}$, called the journey route and represented by $R$, is a walk in $G$, if and only if $\rho(e_i,t_i)=1$ and $t_{i+1}\geq t_i+\zeta(e_i,t_i)$ for all $i<k$. Every journey has a $departure(\mathcal{J})$ and an $arrival(\mathcal{J})$ that refer to journey's starting time $t_1$ and its last time $t_k+\zeta(e_k,t_k)$. 
Journeys are divided into three classes based on their variations based on the temporal and topological distance \cite{xuan2003computing}. Journeys that have earliest arrival times are called  \emph{foremost} {journeys}, journeys with the  smallest topological distance are referred to as \emph{shortest}{ journeys}, 
while the journey that takes the smallest amount of time is called \emph{fastest}. 
Moreover, we call {\em  foremost increasing journey} the ones whose route $\{e_1,e_2, \ldots, e_k\}$ is such that {\em birth-date}$(e_i) \leq$ {\em birth-date}$(e_{i+1})$.

When representing {\em Knowledge-Net} as a TVG ${\cal G}$ we notice that,  due to  zero latency  and to the fact that edges never disappear once created, any shortest journey route in ${\cal G}$ is equivalent to a shortest path on the static graph corresponding to its footprint; moreover, the notion of fastest journey does not have much meaning in this context, because on any route corresponding to a journey, there would be a fastest one.  On the other hand, the notion of foremost journey, and in particular of foremost increasing journey, is extremely relevant as it describes   timely  mobilization flow, i.e., flow that arrives at a node as early as possible.

\subsection{Temporal Betweenness}

Betweenness  is a classic measure of centrality extensively investigated in the context of social network analysis;
the betweenness of a node $v\in V$ in a static graph $G=(V,E)$ is defined as follows:
 
\begin{equation}\label{staticB}
B(v)  =\sum_{u\neq w\neq v\in V}\frac{|P(u,w,v)|}{|P (u,w)|}
\end{equation}

\noindent
where $|P(u,w)|$ is the number of shortest paths from $u$ to $w$ in  $G$, and $|P(u,w,v)|$ is the number of those passing through $v$. Even if static betweenness is ``atemporal", we denote here by $B(v)^ \mathcal{T}$ the static betweenness of a node $v$ in a system whose lifetime is $\mathcal{T}$. 
Typically, vertices with high betweenness centrality direct a greater flow, and thus, have a high load 
placed on them, which is considered as an indicator for their importance as potential gatekeepers in the network.

While betweenness in static graphs is based on the notion of shortest path, its temporal version can be extended into three different measures to  consider shortest, foremost, and fastest journeys for a given lifetime $\mathcal{T}$ \cite{santoro2011time}.
As mentioned earlier, in the context of {\em Knowledge-Net}, fastest betweenness cannot really be defined, and shortest betweenness would coincide with its static counter-part. We, thus, focus on foremost betweenness. Note that the number of foremost journeys between two nodes can be exponential, or even unbounded, and the computation of foremost betweenness is an intractable task. In this paper we consider a  form of foremost betweenness that, although 
still counting possibly an exponential number of journey,  is more manageable.  Foremost betweenness $ TB^ \mathcal{T}_{\cal F} (v)  $  for node $v$ 
with lifetime $\mathcal{T}$ is here defined as follows:

\begin{equation}\label{eq:TemporalBetSantoro}
TB^ \mathcal{T}_{\cal F} (v) =\sum_{u\neq w\neq v\in V}\frac{|\mathcal{F}^\mathcal{T} (u,w,v)|}{|\mathcal{F}^ \mathcal{T}(u,w)|}
\end{equation}

\noindent
where $|\mathcal{F}^ \mathcal{T}(u,w)|$ is the number of foremost  {\em increasing  journey routes} between $u$ and $w$ during time frame $\mathcal{T}$ and $|\mathcal{F}^ \mathcal{T}(u,w,v)|$ is the number of the ones passing through $v$ in the same time frame.  Besides being a little more computationally manageable, choosing   increasing   journey routes emphasizes the role of the first year of  creation   of a connection  in the network.
To take into account possible network disconnections, we multiply the betweenness value $TB^ \mathcal{T}_{\cal F} (v)$
 by the adjustment  coefficient $n(v) \over n$ where $n(v)$ is the number of nodes in the connected component to which $v$ belongs, and $n$ is the total number of nodes. Analogous adjustment is performed for $B(v)$.
 
Highly-ranked vertices for foremost betweenness do not simply act as gatekeepers of flow, like their static counter-part. In fact, they   direct the flow that conveys a message in an {\em earliest} transmission fashion. In other words, intuitively, they  provide some form of ``acceleration" in the flow of information.

\section{Foremost Betweenness of Knowledge-Net}

In this Section we focus on {\em Knowledge-Net}, and we study $TB^ \mathcal{T}_{\cal F} (v)$ for all $v$. Nodes are ranked according to their betweenness values and their ranks are compared with the ones obtained calculating their static betweenness $B^ \mathcal{T}(v)$ in the same time frame. Given the different meaning of those two measures, we expect to see the emergence of different behaviours, and, in particular, we hope to be able to detect nodes with important temporal roles that were left undetected in the static analysis.

\subsection{Foremost Betweenness during the lifetime of the system}

Table \ref{tbl:temporalStaticBet} shows the temporally ranked actors accompanied by their static ranks, and the high ranked static actors with their temporal ranks, both with lifetime $\mathcal{T}=$  [2005-2011]. 
 In our naming convention, an actor  named $Xi(yy)$ is of type $X$, birth date $yy$ and it is indexed by $i$; types are abbreviated as follows: $H$ (human), $L$ (Lab), $A$ (article), $C$ (conference), $J$ (journal), $P$ (project), $C$ (paper citing a publication), $I$ (invited oral presentation), $O$ (oral presentation).
Note that only the nodes whose betweenness has a significant value are considered, in fact betweenness values tend to  lose their importance, especially when the differences in the values of two consecutive ranks are very small \cite{freeman1977set}.

\newcolumntype{g}{>{\columncolor{Gray}}c}
\begin{table}
	\renewcommand{\arraystretch}{1.3}
	\caption{List of   highest   ranked actors according to temporal (resp. static) betweenness,
	accompanied by the corresponding static (resp. temporal) rank in lifetime [2005-2011].}
	\label{tbl:temporalStaticBet}
	\begin{center}
		\centering
		\begin{tabular}{|c|c|c|c|c|c|c|}
			\hline
			\multicolumn{3}{|c|}{Temporal to Static} && \multicolumn{3}{c|}{Static to Temporal}\\
			\cline{1-3}\cline{5-7}
			 Actor & Temporal Rank & Static Rank && Actor &  Static Rank &Temporal Rank \\
			\cline{1-3}\cline{5-7}
			 L1(05)&	1 & 1 &&  L1(05)& 1 &	1\\
			\cline{1-3}\cline{5-7}
			H1(05)	&2  & 	2 &&   H1(05) 	&2 & 2\\
			\cline{1-3}\cline{5-7}
			 A1(06) &		3&  3  &&  A1(06)  &3& 3\\
			\cline{1-3}\cline{5-7}
			 A2(08) &	4& 	4  && 	 A2(08)  &4&	4\\
			\cline{1-3}\cline{5-7}
			 P1(06)	&5 &  	8  && 	  A5(08)  &5   & 12\\
			\cline{1-3}\cline{5-7}
			 A3(07) &		6&  9  &&   A4(09)  &6&	7\\
			\cline{1-3}\cline{5-7}
			 A4(09) &	 7	& 6  &&  P2(08)  &7& 9	\\
			\cline{1-3}\cline{5-7}
			 S1(10)	& 	8&  115 &&  P1(06)  &8&		5\\
			\cline{1-3}\cline{5-7}
			 P2(08) &	 9	 &  7 &&  A3(07)&9&	6\\
			\cline{1-3}\cline{5-7}
			 J1(06) & 10	& 	160  &&   P3(10)  &10& 	17\\
			\cline{1-3}\cline{5-7}
			 C1(07) &11&  	223  &&   A6(11) &11&	18\\
			\cline{1-3}\cline{5-7}
			 A5(08) & 12	&  	5 &&   A8(09) &12&	36\\
			\cline{1-3}\cline{5-7}
			 I1(09) & 	13  &  	28  &&  P4(10) &13&  	22\\
			\cline{1-3}\cline{5-7}
			 O1(05) & 14	 & 	45  &&   P5(11) &14& 	27\\
			\cline{1-3}\cline{5-7}
			 S2(05) & 15	& 	46  &&   H2(05) &15& 	44\\
			\cline{1-3}\cline{5-7}
			 I2(05) & 16	 & 	47  && A7(09) &16&	21\\
			\cline{1-3}\cline{5-7}
			 P3(10) & 	17&  	10  &&  A9(10) &17&	31\\
			\cline{1-3}\cline{5-7}
			 A6(11) & 18	& 	11 &&  P5(11) &18&	69\\
			\cline{1-3}\cline{5-7}
			 C2(10)	  & 19   & 	133  &&   P6(10) &19& 	23\\
			\cline{1-3}\cline{5-7}
			 J2(09)&    20 & 		182  && & &\\
			\cline{1-3}\cline{5-7}
			 A7(09)	& 21  & 	16  && & &\\
			\hline
		\end{tabular}
	\end{center}
\end{table}

Interestingly, the four highest ranked nodes are the same under both measures; in particular, the highest ranked node (L1(05)) corresponds to the main laboratory where the data is collected and it is clearly the most important actor in the network  whether considered in a temporal or in a static way. On the other hand, the table reveals several differences worth exploring. From a first look we see that, while the vertices highest ranked statically appear also among the highest ranked  temporal ones, there are some nodes with insignificant static betweenness, whose temporal betweenness is extremely high. This is the case, for example, of nodes S1(10) and J1(06).  

\subsubsection{The case  of  node  S1(10)}

To provide some interpretation for this behaviour we observe vertex S1(10)  in more details. This vertex corresponds to a poster presentation at a conference in 2010. We explore two insights. First, although S1(10) has a relatively low degree, it has a great variety of temporal connections. Only three out of ten incident edges of S1(10) are connected to actors that are born on and after 2010, and the rest of the neighbours appear in different times, accounting for at least one neighbour appearing each year for which the data is collected. This helps the node to operate as a temporal bridge between different time instances and to perhaps act as a knowledge mobilization accelerator.

Second, S1(10) is close to the centre of the only static community present in [2010-2011] and it is connected to the two most important vertices in the network. The existence of a single dense community, and the proximity to two most productive vertices can explain its negligible static centrality value: while still connecting various vertices S1(10) is not the shortest connector and its betweenness value is thus low. However, a closer temporal look reveals that it plays an important role as an interaction bridge between all the actors that appear in 2010 and later, and the ones that appear earlier than 2010. This role remained invisible in static analysis, and only emerges when we pay attention to the time of appearance of vertices and edges. On the basis of these observations, we can interpret S1(10)'s high temporal betweenness value as providing a fast bridge from vertices created earlier and those appearing later in time. This lends support to the importance of poster presentations that can blend tacit and explicit knowledge mobilization  in human -  poster presentation -  human relations during conferences
 and continue into future mobilization with new non-human actors as was the case for S1(10)  \cite{BB10}.

\subsubsection{The case  of  node  J1(06)}
J1(06), the {\em Journal of Neurochemistry}, behaves similarly to S1(10) with its high temporal and low static rank.
As opposed to S1(10), this node is introduced   very early  in the network (2006); however, it  is only active (i.e. has new incident edges) in 2006 and 2007. 
It has only   three neighbours, A1(06), A3(07), and C1(07), 
all highly ranked vertices statically (A1(06), A3(07)), or temporally (C1(07)).  
Since its neighbouring vertices are directly connected to each other or in close proximity of two hops, J1(06) fails to act as a static short bridge among graph entities. However, its early introduction and proximity to the most prominent knowledge mobilizers helps it become an important temporal player in the network. This is because temporal journeys overlook geodesic distances and are instead concerned with temporal distances for vertices. These observations might explain the high temporal rank of J1(06)  in the knowledge mobilization network.

\subsection{A Finer look at foremost betweenness}

A key question is whether the birth-date of a node is an important factor influencing its temporal betweenness. To gain insights, we conducted a finer temporal analysis by considering $TB^ \mathcal{T}_{\cal F}$ for all possible birth-dates, i.e, for $ \mathcal{T}=$ [x,2011],  $\forall x\in$
$ \{2005,2006,2007,2008,2009,2010,$
$2011\}$. This allowed us to observe how temporal betweenness varies depending on the considered birth-date.

Before concentrating on selected vertices (statically or temporally important with at least one interval), and analysing them in more detail, we briefly describe a temporal community detection mechanism that we employ in analysis.

\subsubsection{Detection of temporal communities}

\begin{figure}
	\centering
	\includegraphics[width=0.5\linewidth]{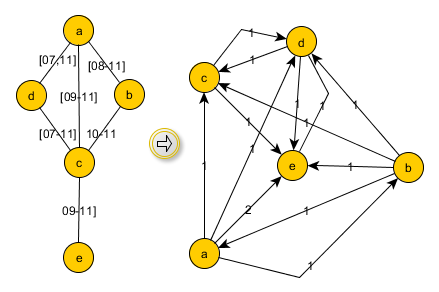}
	\caption{Transformation of a temporal graph into a weighted graph used for	community detection. 
	}
	\label{fig:TemporalCommunities}
\end{figure}

We approximately detect communities existing in temporal networks. To detect communities involving $x$, we first determine the temporal foremost journeys arriving at or leaving from $x$. We then replace each journey with a single edge, creating a static graph with an edge between $x$ and all the vertices that are reachable from or can reach $x$ in a foremost manner. For instance, Fig. \ref{fig:TemporalCommunities} shows the transformation of a graph into a directed weighted graph that is used for community detection. We finally apply existing directed weighted community detection algorithms to compute communities around $x$ \cite{Gomez09}. 
The model is an approximation since it overlooks the role that is played in communities by vertices that fall along  journeys while not being their start or end-points; however,   it is sufficient for our purposes to give an indication of the community formation around  a node.

\subsubsection{The case of node P1(06)}

This is a research project led by the principle investigator at L1(05). The project was launched in 2006 and its official institutional and funded elements wrapped-up in 2011. Data in Table \ref{tbl:temporalStaticBet} support that P1(06) has similar temporal and static ranks with regards to its betweenness in lifetime [2005-2011]. One could conclude that the temporal element does not provide additional information on its importance and that the edges that are incident to P(06)-1 convey the same temporal and static flow. However, there is still an unanswered question on whether or not edges act similarly if we start observing the system at different times. Will a vertex keep its importance throughout the system's lifetime?

\begin{figure}
	\centering
	\includegraphics[width=0.7\linewidth]{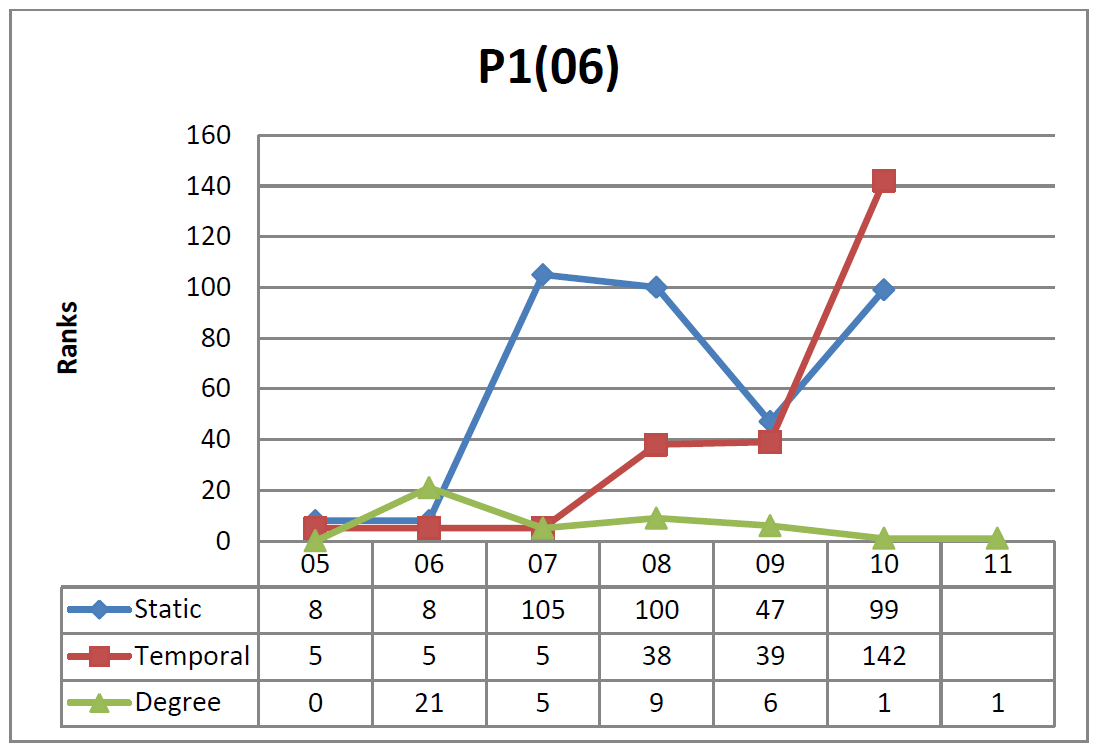}
	\caption{Comparison between different values for vertex P1(06).  Ranks of the vertex in the last interval
are  not provided as both betweenness values are zero.}
	\label{fig:RP38_2006-2011}
\end{figure}

The result of such analysis is provided in  Fig. \ref{fig:RP38_2006-2011}, where  $TB^ \mathcal{T}_{\cal F} (P1(06))$  is calculated for each birth-date (indicated in the horizontal axis), with all intervals ending in 2011.

While both equally important during the entire lifetime [2005-2011] of the study, this project seems to assume a rather more relevant temporal role when observing the system in a lifetime starting in year 2007  (i.e., $\mathcal{T} = $[2007-2011]), when its static betweenness is instead negligible. This seems to indicate that the temporal flow of edges incident to P1(06) appearing from 2007 on is more significant than the flow of the edges that appeared previously.

With further analysis of P1(06)'s neighbourhood in [2007-2011], we can formulate technical explanations for this behaviour. First, its direct neighbours also have better temporal betweenness than static betweenness. Moreover, its neighbours belong to various communities, both temporally and statically. However, looking at the graph statically, we see several additional shortest paths that do not pass through P1(06)  (thus making it less important in connecting those communities). In contrast, looking at the graph temporally P1(06) acts as a mediator and accelerator between communities.
More specifically, we observe that the connections P1(06) creates in 2006 contribute to the merge of different communities that appear only in 2007 and later. When observing within interval [2006-2011], we then see that P1(06) is quite central from a static point of view, because the appearance of time of edges does not matter but, when observing it in lifetime [2007-2011] node P1(06) loses this role and becomes statically peripheral because the newer connections relay information in an efficient temporal manner.

In other words, it seems that P1(06) has an important role for knowledge acceleration in the period 2007-2011, a role that was hidden in the static analysis and that does not emerge even from an analysis of consecutive static snapshots. For research funders, revealing a research project's potentially invisible mobilization capacity is relevant. Research projects can thus be understood beyond mobilization outputs and more in terms of networked temporal bridges to broader impact.

\subsubsection{The case of node A3(07)}   
The conditions for A3(07), a paper published in 2007, illustrate a different temporal phenomenon. Node A3(07) has several incident edges in 2007 (similarly to node P1(06)) when both betweenness measures are high. Peering deeper into the temporal communities formed around A3(07) is revealing: up to 2007, this vertex is two degrees from vertices that connect two different communities in the static graph. The situation radically changes however with the arrival of edges in 2008 that modify the structure of those communities and push A3(07) to the periphery. The shift is dramatic from a temporal perspective because A3(07) loses it accelerator role where its temporal betweenness becomes negligible, while statically there is only a slight decrease in betweenness. The reason for a dampened decrease in static betweenness is that this vertex is close to the centre of the static community, connecting peripheral vertices to the most central nodes of the network (such as L1(05) and H1(05)). It is mainly proximity to these important vertices that sustains A3(07)'s static centrality.

Such temporal insights lend further support to understanding mobilization through a network lens coupled with sensitivity to time. A temporal shift to the periphery for an actor translates into decreased potential for sustained mobilization.

\begin{figure}
\centering
\includegraphics[width=0.7\linewidth]{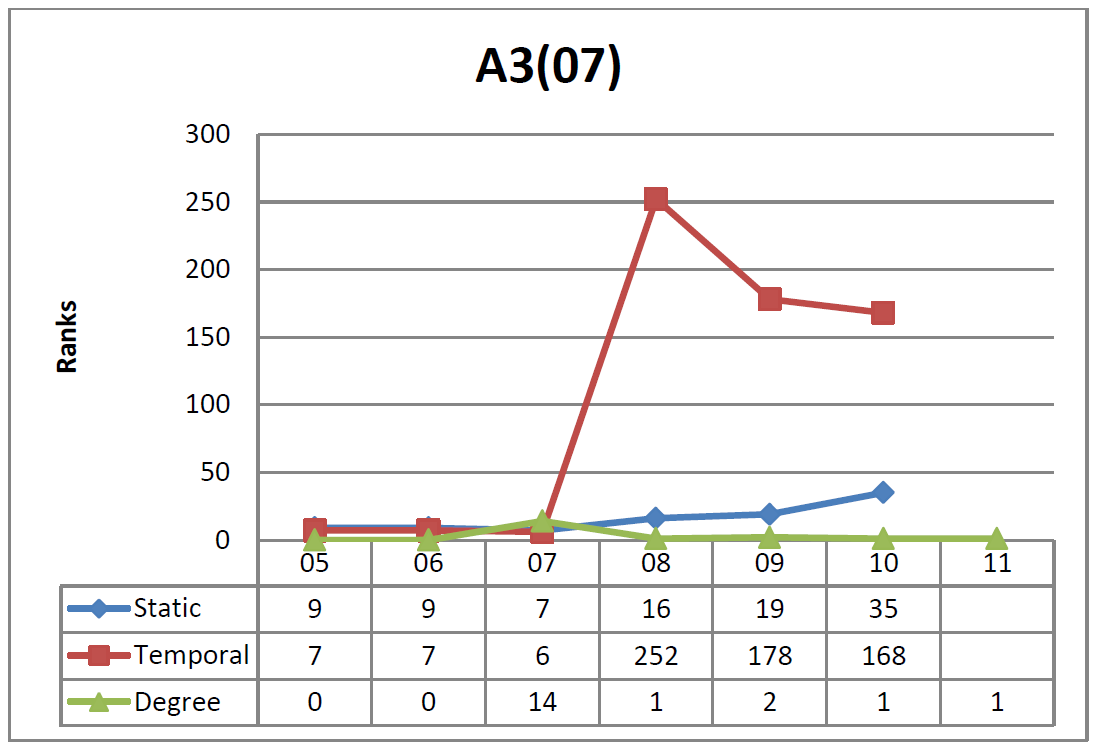}
\caption{Comparison between different values for vertex A3(07). Ranks of the vertex in the last interval
are  not provided as both betweenness values are zero.} 
\label{fig:2007P01}
\end{figure}
 
\section{Invisible  Rapids and Brooks}

On the basis of our observations, we define two concepts to differentiate the static and temporal flow of vertices in  Knowledge Mobilization networks. 
 We call {\em rapids}  the nodes with high foremost betweenness, meaning that they can potentially mobilize knowledge in a timelier manner;
 and {\em brooks} the ones with insignificant  foremost betweenness. 
 Moreover,
 we call {\em invisible  rapids} those vertices whose temporal betweenness rank  is considerably more significant than their static rank 
(i.e., the ones whose centrality was undetected by  static betweenness),
  and
  {\em invisible brooks} the ones whose static betweenness is considerably higher than their temporal betweenness, 
  meaning that these vertices can potentially be effective knowledge mobilizers, yet they are not acting as effectively as others due to slow or non-timely relations.  

Invisible rapids and brooks can be present in different lifetimes as their temporal
 role might be restricted to some time intervals only; 
for example, as we have seen in the previous Section, 
 S1(10) and J1(06) are invisible  rapids in  ${\cal T} = $ [2005-2011],
 P1(06) is an invisible rapid in ${\cal T} = $ [2007-2011],  A3(07) is an invisible brook in   ${\cal T} = $ [2008-2011].

Tables \ref{tbl:rapids} and \ref{tbl:bottlenecks} indicate the major invisible rapids and brooks   observed in  {\em Knowledge-Net}.

 \begin{table}[tbh]
	\renewcommand{\arraystretch}{1.3}
	\caption{Major invisible rapids} \label{tbl:rapids}
		\centering
		\begin{tabular} {|c|c|c|c|c|}
			\hline
			 Actor & Time  & Temp. Rank & Stat. Rank & Type \\
			  \hline
	                 P1(06) & [07-11] &	5 &	105 & project  \\
			\hline
			S1(10)& [05-11] &	8 &	115 & poster  \\
				 & [06-11] &	8 &	113 &   \\
				  & [07-11] &	7 &	115 &   \\
				   & [08-11] &	5 &	104 &   \\
			                                                           \hline
		          J1(06) & [05-11] &	10 &	160 & journal \\
		          	 & [06-11] &	10 &	154 &   \\
			         & [07-11] &	10 &	223 &   \\
			           \hline
		      		          C1(07) & [05-11] &	11 &	223 & citing publication \\
		          	 & [06-11] &	11 &	220 &   \\
					    \hline
			   J2(09)    & [06-11] &	17 &	179 & journal \\
			                  & [07-11] &	16 &	182 &   \\
			                  \hline
			                      C2(10) & [05-11] &	19 &	133 &  citing poster  \\
		          	 & [06-11] &	16 &	132 &   \\
			 	 & [07-11] &	15 &	133 &   \\

			        \hline

		\end{tabular}
\end{table}

The presence of a poster presentation, a research project, two journals and a conference publication among the invisible  rapids supports that different types of mobilization actors can impact timely mobilization while not being as effective at creating short paths among entities for knowledge mobilization. In other words, they can play a role of accelerating knowledge mobilization, but to a concentrated group of actors.

 \begin{table}[tbh]
	\renewcommand{\arraystretch}{1.3}
	\caption{Major invisible brooks}  \label{tbl:bottlenecks}
		\centering
		\begin{tabular} {|c|c|c|c|c|}
			\hline
			 Actor & Time  & Stat. Rank & Temp. Rank & Type \\
			\hline
			J3(08) & [08-11] &	9 &	117 & journal  \\
				      & [09-11] &	12 &	84 &   \\
			                                                           \hline
		         C3(11)& [08-11] &	10 &	191 & citing publication \\
		          	               & [09-11] &	15 &	153 &   \\
			               		\hline
					         
		      C4(11)  & [08-11] &	15 &	105 & citing publication \\
\hline
                        H2(05) & [06-11] &	16 &	118 & researcher \\
		          	               & [07-11] &	15 &	134 &   \\
			               		\hline
                                 A3(07)
                                  & [08-11] &	16 &	187 & publication \\
                        			        \hline
		 C5(07)   & [08-11]  & 18 & 158 & citing publication \\
 \hline
		\end{tabular}
\end{table}

As for invisible  brooks,   
we observe a journal (the {\em Biochemica et Biophysica Acta-Molecular Cell Research} (J3(08)), three papers (C3(11), C4(07), and C5(07))
 that cite publications by the main laboratory in the study (L1(05)),  a publication (A3(07)) mobilizing knowledge from members of L1(05), and 
 a research assistant who worked on several research projects as an HQP.
In comparison with invisible rapids,  there  is a wider variety in the type of mobilization actors that act as brooks which does not readily lend itself to generalization.

Interestingly, we see the presence of journals among invisible rapids and brooks. From our analysis, it seems that journals can hold strikingly opposite roles: on the one hand they can contribute considerably to more timely mobilization of knowledge while not being very strong bridges between communities; while on the other hand, they can play critical roles in bridging network communities, but at a slow pace. 
A brook, the journal {\em Biochemica et Biophysica Acta-Molecular Cell Research} (J3(08)), for
example, helped mobilize knowledge in two papers for L1(05) (in 2008 and 2009) and is a journal in which a paper (in 2011) citing a L1(05) publication 
was also published. 
Given expected variability in potential mobilization for a journal, it is not surprising to see these mobilization actors at both ends of the spectrum. 

In contrast, the presence of a research project as an invisible  rapid is meaningful. It is meaningful in two ways. First, because when public funders invest in research projects as mobilization actor, an implicit if not explicit measure of success is timely mobilization Ð with potential impact inside and outside of academia \cite{gaudet2014PP}. Ranking as a rapid (for a mobilization actor) is one measure that could therefore help funding agencies monitor and detect temporal change in mobilization networks. 
Second, a research project as rapid is meaningful because by its very nature a research project can help accelerate mobilization for the full range of mobilization actors, including other research projects. As such, it is not surprising that they can become temporal conduits to knowledge mobilization in all of its forms.

\section{Conclusion}

In this paper, we proposed the use of a temporal betweenness measure (foremost betweenness)  to analyse a knowledge mobilization network that had been already studied using classical ``static" parameters. Our goal was to see the impact on the perceived static central nodes when employing a measure that explicitly takes time into account. We observed interesting differences. In particular, we witnessed the emergence of invisible rapids: nodes whose static centrality was considered negligible, but whose temporal centrality appears relevant. Our interpretation is that nodes with high temporal betweenness contribute to accelerate mobilization flow in the network and, as such, they can remain undetected when the analysis is performed statically. 
We conclude that foremost betweenness is  a crucial tool to understand the temporal role of the actors in a dynamic network, and that 
the combination of static and temporal betweenness is complementary to provide insights into their importance and centrality.

Temporal network analysis as performed here is especially pertinent for KM research that must take time into account to understand academic research impact beyond the narrow short-term context of academia. Measures of temporal betweenness, as studied in this paper, can provide researchers and funders with critical tools to more confidently investigate the role of specific mobilization actors for short and long-term impact within and beyond academia.  The same type of analysis
 could clearly be beneficial when applied to any other temporal context.

In conclusion, we focused here on a form of temporal betweenness designed to detect accelerators. This is only a first step towards understanding temporal dimensions of social networks; other measures are already under investigation.

\section*{Acknowledgment}

This work was partially supported by a NSERC Discovery Grant and by Dr. Flocchini's University Research Chair.

\bibliographystyle{plain}

\begin{thebibliography}{10}

\bibitem{ACPQS11}
F. Amblard, A. Casteigts, P. Flocchini, W.  Quattrociocchi, and N. Santoro.
\newblock On the temporal analysis of scientific network evolution. 
\newblock {\em International Conference on Computational Aspects of Social Networks} (CASoN), 169-174, 2011.



\bibitem{BB10}
S. Beaudoin, X. Roucou X.
\newblock Genetic prion mutants inhibit two RNA granules, P-Bodies and stress granules. 
\newblock {\em PrPCANADA}, Ottawa, Canada, 2010.


\bibitem{Binz2014}
C.  Binz, B. Truffer, and L. Coenen.
\newblock {Why space matters in technological innovation systemsâ
Mapping  global knowledge dynamics of membrane bioreactor technology}.
\newblock {\em Research Policy}, 43(1):138--155, February 2014.


\bibitem{xuan2003computing}
B~Bui Xuan, A. Ferreira, and A.  Jarry.
\newblock {Computing shortest, fastest, and foremost journeys in dynamic
  networks}.
\newblock {\em International Journal of Foundations of Computer Science},
  14(02):267--285, 2003.
  
\bibitem{Boland2012}
W.~P. Boland, P.~W.B. Phillips, C.~D. Ryan, and S.
  McPhee-Knowles.
\newblock {Collaboration and the Generation of New Knowledge in Networked
  Innovation Systems: A Bibliometric Analysis}.
\newblock {\em Procedia - Social and Behavioral Sciences}, 52:15--24,  
  2012.

 \bibitem{CFMS10}
A.  Casteigts, P. Flocchini, B. Mans, and N.  Santoro.
 \newblock {Deterministic Computations in Time-Varying Graphs: Broadcasting under Unstructured Mobility}.
\newblock  {\em Proc. 6th IFIP conference on Theoretical Computer Science},  111--124,  2010.


\bibitem{CFMS12}
A.  Casteigts, P. Flocchini, B. Mans, and N.  Santoro.
 \newblock {Measuring Temporal Lags in Delay-Tolerant Networks}.
\newblock {\em IEEE Transactions on Computers},  63(2), 397--410, 2014.


 

\bibitem{CFQS12}
A.  Casteigts, P. Flocchini, W. Quattrociocchi, and N.  Santoro.
 \newblock {Time-varying graphs and dynamic networks}.
\newblock {\em International Journal of Parallel, Emergent and Distributed
  Systems}, 27(5):387--408, 2012.

 
\bibitem{chan2006synergy}
K. Chan and J. Liebowitz.
\newblock {The synergy of social network analysis and knowledge mapping: a case
  study}.
\newblock {\em International Journal of Management and Decision Making},
  7(1):19--35, 2006.

\bibitem{Eppler2001}
M.J. Eppler.
\newblock {Making knowledge visible through intranet knowledge maps: concepts,
  elements, cases}.
\newblock   {\em  Proc.  34th Annual Hawaii International
  Conference on System Sciences},   I  2001.

\bibitem{freeman1977set}
L.~C Freeman.
\newblock {A set of measures of centrality based on betweenness}.
\newblock {\em Sociometry},   35--41, 1977.

  \bibitem{GVOM13}
A. Galati, V.   Vukadinovic, M.  Olivares, and S.  Mangold.
\newblock {Analyzing temporal metrics of public transportation for designing scalable delay-tolerant networks},
\newblock   {\em Proc.   8th ACM Workshop on Performance Monitoring and Measurement of Heterogeneous Wireless and Wired Networks},
37-44, 2013 
 
\bibitem{gaudet2013takes}
J. Gaudet.
\newblock {It takes two to tango: knowledge mobilization and ignorance
  mobilization in science research and innovation}.
\newblock {\em Prometheus}, 13(3):169-187, 2013.

\bibitem{gaudet2014PP}
J. Gaudet.
\newblock {The Mobilization-Network Approach for the Social Network Analysis of
  Knowledge Mobilization in Science Research and Innovation}.
\newblock {\em uO Research}, (PrePrint):1--28, 2014.


\bibitem{Gomez09}
S. G{\'o}mez,   P. Jensen, and A. Arenas.
\newblock {Analysis of community structure in networks of correlated data}.
\newblock {\em Physical Review E},  80(1), 2009.


\bibitem{JLW07}
E.P.C. Jones, L. Li, J.K. Schmidtke, and P.A.S.  Ward.
\newblock {Practical routing in delay-tolerant networks}.
\newblock {\em IEEE Transactions on Mobile Computing},  6(8), 943--959, 2007.

 \bibitem{KosKW08}
 G. Kossinets, J. Kleinberg,   and D. Watts.
\newblock {The structure of information pathways in a social communication network},
\newblock {\em Proc.  14th International Conference on Knowledge Discovery and Data Mining} (KDD), 435--443, 2008.
 

 
\bibitem{Kostakos09}
V. Kostakos.
\newblock {Temporal graphs}.
\newblock {\em Physica A},  388(6), 1007--1023, 2009.

 
\bibitem{Kim2012}
H. Kim and R. Anderson.
\newblock {Temporal node centrality in complex networks}.
\newblock {\em Physical Review E}, 85(2):026107,  2012.

\bibitem{klenk2010quantifying}
N.~L Klenk, A. Dabros, and G.~M Hickey.
\newblock {Quantifying the research impact of the Sustainable Forest Management
  Network in the social sciences: a bibliometric study}.
\newblock {\em Canadian Journal of Forest Research}, 40(11):2248--2255, 2010.

\bibitem{KLHC13}
M. Konschake, H.~HK Lentz, F.~J Conraths, P. H{\"o}vel, and
  Thomas Selhorst.
\newblock On the robustness of in-and out-components in a temporal network.
\newblock {\em PloS one}, 8(2), 2013.

\bibitem{kossinets2006empirical}
G. Kossinets and D.~J. Watts.
\newblock {Empirical analysis of an evolving social network}.
\newblock {\em Science}, 311(5757):88--90, 2006.



\bibitem{KLO10}
F. Kuhn, N.Lynch and R. Oshman.
\newblock {Distributed computation in dynamic networks},
\newblock {\em Proc.   42nd ACM Symposium on Theory of Computing} (STOC), 513--522, 2010.
 

\bibitem{LSS13}
H. H.~K. Lentz, T. Selhorst, and I.~M. Sokolov.
\newblock Unfolding accessibility provides a macroscopic approach to temporal
  networks.
\newblock {\em Phys. Rev. Lett.}, 110:118701--118706, 2013.



\bibitem{LW09b}
C. Liu,  and  J. Wu.
\newblock {Scalable Routing in Cyclic Mobile Networks}.
\newblock {\em IEEE Transactions on Parallel and Distributed Systems},  20(9),1325--1338,  2009.

 
\bibitem{MiCS14}
O. Michail, I.  Chatzigiannakis,  and P. Spirakis.
\newblock {Distributed computation in dynamic networks},
\newblock {\em Journal of Parallel and Distributed Computing}, 74(1), 2016--2026, 2014.  


\bibitem{MBA12}
A.~Y. Mutlu, E. Bernat, and S. Aviyente.
\newblock A signal-processing-based approach to time-varying graph analysis for
  dynamic brain network identification.
\newblock {\em Computational and mathematical methods in medicine}, 2012.


\bibitem{QCL11}
W. Quattrociocchi, R. Conte, and E. Lodi.
\newblock Opinions manipulation: Media, power and gossip.
\newblock {\em Advances in Complex Systems}, 14(4):567--586, 2011.


\bibitem{SVMM14}
H. Saba, V.~C Vale, M.~A Moret, and J-G  Miranda.
\newblock Spatio-temporal correlation networks of dengue in the state of bahia.
\newblock {\em BMC Public Health}, 14(1):1085, 2014.

\bibitem{santoro2011time}
N.  Santoro, W.  Quattrociocchi, P. Flocchini, A.  Casteigts, and   Fr\'{e}d\'{e}ric Amblard.
\newblock {Time-Varying Graphs and Social Network Analysis: Temporal Indicators
  and Metrics}.
\newblock   {\em Proc. of 3rd AISB Social Networks and Multiagent Systems
  Symposium (SNAMAS)},   32--38, 2011.

\bibitem{HS12}
J.~Saramaki P.~Holme.
\newblock {Temporal networks}.
\newblock {\em Physics reports}, 519(3):97--125, 2012.


\bibitem{Tang2009}
J. Tang,  M. Musolesi, C. Mascolo, and V. Latora.
\newblock {Temporal distance metrics for social network analysis}.
\newblock  {\em  Proc.  2nd ACM Workshop on Online Social Networks} (WOSN),   31-36,  2009.

 \bibitem{TBK07}
C. Tantipathananandh, T. Berger-Wolf, D. Kempe.
\newblock {A framework for community identification in dynamic social networks},
\newblock {\em Proc.     13th ACM SIGKDD International Conference on Knowledge Discovery and Data Mining},
717-726, 2007 
 
 \end{thebibliography}

\end{document}